\documentstyle[pra,aps]{revtex}
          \begin{document}
          \draft
          \title{Quantum Robots and Environments} 
          \author{Paul Benioff\\
           Physics Division, Argonne National Laboratory \\
           Argonne, IL 60439 \\
           e-mail: pbenioff@anl.gov}
           \date{\today}

          \maketitle
          \begin{abstract}  
Quantum robots and their interactions with environments of
quantum systems are described and their study justified.  A
quantum robot is a mobile quantum system that includes an on
board quantum computer and needed ancillary systems.  Quantum
robots carry out tasks whose goals include specified changes in
the state of the environment or carrying out measurements on the
environment. Each task is a sequence of alternating computation
and action phases.  Computation phase activites include 
determination of the action to be carried out in the next phase and
recording of information on neighborhood environmental system states. 
Action phase activities include motion of the quantum robot and changes 
in the neighborhood environment system states.  Models
of quantum robots and their interactions with environments are
described using discrete space and time. To
each task is associated a unitary step operator $T$ that gives
the single time step dynamics. $T=T_{a}+T_{c}$ is a sum of action
phase and computation phase step operators.  Conditions that
$T_{a}$ and $T_{c}$ should satisfy are given along with a
description of the evolution as a sum over paths of completed
phase input and output states.  A simple example of a task
carrying out a measurement on a very simple environment is
analyzed in detail.  A decision tree for the task is presented
and discussed in terms of the sums over phase paths. It is seen
that no definite times or durations are associated with the
phase steps in the tree and that the tree describes the successive
phase  steps in each path in the sum over phase paths. 
 \end{abstract}

\section{Introduction}

Much of the impetus to study quantum computation, either as 
networks of quantum gates \cite{Vedraletal,Barencoetal} (See
\cite{EkJo} for a review) or as Quantum Turing Machines
\cite{Benioff8082,Benioff86,Deutsch85,Deutsch89,BenioffQBE}, is
based on the increased efficiency of quantum computers compared
to classical computers for solving some important problems
\cite{Shor,Grover}. Realization of this goal or use of quantum
computers to simulate other physical systems
\cite{Feynman82,Deutsch85} requires the eventual physical
construction of quantum computers.  However, as emphasized
repeatedly by Landauer \cite{Landauer}, there are serious
obstacles to such a physical realization. 

In much of the work done so far quantum computers are considered to be
free standing systems.  Interactions with external environmental
systems are to be avoided either by use of error correcting codes
\cite{Lf} or other methods of making resilient quantum computers
\cite{KnLfZu}. However one can take a different view by
considering quantum computers to be parts of larger systems where 
interactions between quantum computers and systems
external to the quantum computer are an essential part of the
overall system dynamics. They are not something to be avoided or
minimized.  

This view will be followed here by consideration of quantum robots and their
interactions with environments of quantum systems. A
quantum robot is considered to be a mobile system with a quantum
computer and needed ancillary systems on board.  The quantum
robot moves in and interacts with an external environment of
quantum systems.

There are also foundational aspects that justify the study of
quantum computers and of  quantum robots interacting with
environments.  These are based on the fact that validation of a
physical theory such as quantum mechanics involves comparison of
numerical values calculated from theory with experimental
results.  If quantum mechanics is universally valid (and there is
no reason to assume otherwise), then both the systems that carry
out theoretical calculations and the systems that carry out
experiments must be described within quantum mechanics.  It
follows that the systems that test the validity of quantum
mechanics must be described by the same theory whose validity
they are testing.  That is quantum mechanics must describe its
own validation to the maximum extent possible \cite{PerZur}.

Because of these self referential aspects, limitations in
mathematical systems expressed by the G\"{o}del theorems lead one
to expect that there may be interesting questions of self
consistency and limitations in such a description.  Limitations
on self observation by quantum automata
\cite{Albert,Breuer,Peres} may also play a role here.

Investigation of these questions for quantum mechanics requires
that one have well defined completely quantum mechanical
descriptions of systems that compute theoretical values and of
systems that carry out experiments.  So far there has been much
work on quantum computers.  These are systems that can, in
principle at least, carry out computation of theoretical values
for comparison with experiment.  However there has been no
comparable development of a quantum mechanical description of
robots.  These are systems that can, in principle at least, carry
out experiments.

Another related reason that supports study of quantum robots is
that they provide a {\em very small} first step towards a quantum
mechanical description of systems that are aware of their
environment, make decisions, are intelligent, and create theories
such as quantum mechanics \cite{Penrose,Stapp,Squires}. If
quantum mechanics is universal, then these systems must also be
described in quantum mechanics to the maximum extent possible.

From the foundational point of view, the main point of this paper
is that quantum robots and their interactions with environments
provide a well defined platform for investigation of many
interesting questions generated by the above considerations.  For
example one can investigate if the approach taken here is useful,
and, if not, how the definitions and platform need to be changed. 
However without a well defined basis one cannot hope to make
progress.

The next section provides more details on the description of
quantum robots and their interactions with environments. The
dynamics of the interactions of quantum robots with environments
is described in terms of tasks to be implemented by the quantum
robot. Tasks are described as alternating sequences of
computation and action phases with the goal of either making
specified changes in the state of the environment or carrying out
measurements on the environment. Examples of tasks are given. It
is also noted that the description given of a task makes no
explicit use of a quantum computer.  This raises the question if
it is sufficient to limit consideration to special purpose
dedicated quantum robots without on board computers. A suggested
negative answer, based on efficiency and universality, is given
to support the need for on board quantum computers.

Section \ref{SMQRE} provides a specific model of the dynamics of
quantum robots and their interactions with environments. The
model includes simplifying assumptions of discrete time and
space. Properties of the unitary time step operator $T$
associated with each task for a quantum robot are described in
terms of properties of the action phase ($T_{a}$) and computation
phase ($T_{c}$) step operators where $T=T_{a}+T_{c}$. 

In section \ref{SOP} the evolution of the overall system state given by
$\Psi(n) =T^{n}\Psi(0)$ is organized into a sum over phase paths. 
This is a sum over variable length paths  of input and output
states of successive completed phases of a task. Each path
includes a sum over all distributions of steps within each phase
subject to the total number of steps equaling $n$. The completion
and initiation of each phase in a task are regulated by an on
board control qubit.

A very simple example of a task for a quantum robot in a very
simple environment consisting of one particle on a 1-D lattice 
is analyzed in detail in Section \ref{AVSE}.  The task consists
of measuring the distance between the quantum robot and the
particle by stepwise motion of the quantum robot to the particle,
recording the numbor of steps needed, and returning the quantum
robot to its original position. For each initial position of the
quantum robot and particle the phase path sum contains just one
path. The sums over initial path segment lengths and
distributions of individual phase durations remain.  An action
and computation phase decision tree for the task is described.

The material presented so far is discussed in Section \ref{D}. It
is noted that, since the example decision tree applies to quantum
mechanical processes, no definite duration or completion  times
are associated with the steps in the tree. However, the time
ordering of the steps in the tree is preserved.  If the phase
path sum contains more than one path, because the initial state
is a linear sum of different robot and particle position states
or $T$ contains errors, then the decision tree applies to each
phase path in the sum. 

The paper concludes with a reemphasis of the need for a well
defined platform for discussion of properties of quantum systems
that make computations and carry out experiments and are
intelligent. Also the speculative possibility of a Church Turing
type hypothesis for the class of physical experiments is noted.

It must be emphasized that the language used in this paper to
describe quantum robots and their interactions with environments
is carefully chosen to avoid any suggestions that these systems
are aware of their environment, make decisions, carry out
experiments or make measurements, or have other properties
characteristic of intelligent or conscious systems.   The quantum
robots described here have no awareness of their environment and
do not make decisions or measurements.  They are inanimate
physical systems that differ in detail only from other physical
systems such as atoms or any other quantum systems.

Some aspects of the ideas presented here have already occurred in
earlier work.  Physical operations have been described as
instructions for well-defined realizable and reproducible
procedures \cite{FouRan}, and quantum state preparation and
observation procedures have been described as instruction
booklets or programs for robots \cite{BenEks}.  However these
concepts were not described in detail and the possibility of
describing these procedures or operations quantum mechanically
was not mentioned. Also quantum computers had not yet been
described. 

More recently use of the electronic states of ions in a linear
ion trap as an apparatus (and a quantum computer register) to
measure properties of vibrational states of the ions has been described
\cite{HeMi}. Also quantum mechanical Maxwell's demons \cite{Lloyd} and
oracle quantum computing \cite{BBBV,BBCM} can be considered as the interaction 
of a quantum computer with an external environment in order to learn something
about the external system.  The same holds for Grover's \cite{Grover}
algorithm where the data base can be considered as a system external to the
quantum computer \cite{ref}. Quantum robots and their interactions with
environments were also discussed earlier by the author \cite{BenSFA}.
However much of the discussion was limited to environments consisting of
quantum registers.

Interactions between the environment and systems are also considered in 
other work on environmentally induced
superselection rules \cite{Zur,Joo}. Here emphasis is on interactions
between the environment and a system as a measurement apparatus that
stabilize a selected basis (the pointer basis) of states of the
apparatus.

\section{Quantum Robots}
\label{QR}

As noted quantum robots are considered here to be mobile systems
that have a quantum computer and any other needed ancillary
systems on board.  Quantum robots move in and interact (locally)
with environments of quantum systems.  Since quantum robots are
mobile, they are limited to be quantum systems with finite
numbers of degrees of freedom.

The on board quantum computer can be described as a quantum
Turing machine, a network of quantum gates, or any other suitable
model.  If it is a quantum Turing machine, it consists of a
finite state head moving on a finite lattice of qubits.  The
lattice can have distinct ends.  However it seems preferable if
the lattice is closed (i.e. cyclic).  If the computer is a
network of quantum gates then it should be a cyclic network with
many closed internal quantum wire loops and a limited number of
open input and output quantum wires.  Even though acyclic
networks are sufficient for the purposes of quantum computation
\cite{Yao} cyclic ones are preferable for quantum robots.  One
reason is that interactions between these networks and the
environment are simpler to describe and understand than those
containing a large number of input and output lines. Also the
only known examples of {\em very} complex systems that are aware
of their environment and are presumably intelligent, contain
large numbers of internal loops and internal memory storage. 

Environments consist of arbitrary numbers and type of systems
moving in 1-, 2-, or 3-dimensional spatial lattices.  This is
based on the simplifying assumption for this paper that space and
time are discrete.  The component systems can have spin or other
internal quantum numbers and can interact with one another or be
free.  Environments can be open or closed.  If they are open then
there may be systems that remain for all time outside the domain
of interaction with the quantum robot that can interact with and
establish correlations with other environment systems in the
domain on the robot.  

The dynamics of a quantum robot and its interactions with the
environment is described here in terms of {\em tasks}.  Tasks can be
described by their goals, or desired results of carrying out the tasks, and
their dynamics, or the types of steps carried out to arrive at the goal. 
Goals of tasks include the carrying out of desired  changes in the state
of the environment and the carrying out of measurements by transfer of
information from the environment to the quantum robot. Tasks of the first
type (with a goal of a desired environment state change) are similar to the 
computation of functions with a quantum computer with the goal 
being the carrying out of a specified function computation.

An example of this type of task is "move each system in region
$R$ 3 sites to the right if and only if the destination site is
unoccupied."  Implementation requires specification of a path to
be taken by the quantum robot in executing the task. Some method
of determining when it is inside or outside of the specified
region and making appropriate movements must be available. In
this case if there are $n$ systems in region $R$ at locations
$x_{1},x_{2}, \cdots ,x_{n}$ in region $R$ then the initial state
of the regional environment, $ \vert \underline{x}\rangle =
\otimes_{j=1}^{n}\vert x_{j} \rangle$  becomes
$\otimes_{j=1}^{n}\vert x_{j}+3 \rangle = \vert
\underline{x+3}\rangle$ provided all destination sites are
unoccupied.   

If the initial state of the regional environment is a linear
superposition of states $\psi
=\sum_{\underline{x}}c_{\underline{x}}\vert \underline{x}\rangle$
of n-system position states $\vert \underline{x}\rangle$ in $R$
then the final state of the regional environment is given by
$\sum_{\underline{x}}c_{\underline{x}}\vert
\underline{x+3}\rangle$. Correlations between the initial
configuration states $\vert \underline{x}\rangle$ and final
states $\theta_{\underline{x}}$ of the quantum robot may be
introduced by carrying out the task.  However this is not
necessary, in principle at least, because the task is reversible.

The above description shows that quantum robots can carry out the
same task on many different environments simultaneously.  This
can be done by use of an initial state of the quantum robot plus
environment that is a linear superposition of different
environment basis states.  For quantum computers the
corresponding property of carrying out many computations in
parallel has been known for some time \cite{Deutsch85}.  Whether
the  speedup provided by this parallel tasking ability can be
preserved for some tasks, as is the case for Shor's \cite{Shor}
or Grover's algorithms \cite{Grover} for quantum computers,
remains to be seen.

There are also many tasks that are irreversible. An example is
the task "clean up the region $R$ of the environment" where
"clean up" has some specific description such as "move all
systems in $R$ to some fixed pattern".  This task is irreversible
because many initial states of systems in $R$ are taken into the
same final state.  It can be made reversible by storing somewhere
in the environment outside of $R$ a copy of each component of the
initial state of the systems in $R$.  For example if $\psi
=\sum_{\underline{x}}c_{\underline{x}}\vert \underline{x}\rangle$
is the initial state, then the copy operation is given by
$\sum_{\underline{x}}c_{\underline{x}}\vert \underline{x}\rangle
\vert \underline{0}\rangle_{cp} \longrightarrow
\sum_{\underline{x}}c_{\underline{x}}\vert \underline{x}\rangle
\vert \underline{x}\rangle_{cp}$ where $\vert
\underline{x}\rangle_{cp}$ is the copy state.

This operation of copying relative to the states in some basis
avoids the limitations imposed by the no-cloning theorem
\cite{WooZur} because an unknown state $\psi$ is not being
copied.  The price paid is that copying relative to some basis
introduces branching into the process in that correlations are
introduced between the state of systems in the copy region and
states of systems in $R$. This is the quantum mechanical
equivalent of the classical case of making a calculation of a
many-one function reversible by copying and storing the input
\cite{Bennett}.

In the above case carrying out the cleanup on the state
$\sum_{\underline{x}}c_{\underline{x}}\vert \underline{x}\rangle
\vert \underline{x}\rangle_{cp}$ corresponds to the operation
$\sum_{\underline{x}}c_{\underline{x}}\vert \underline{x}\rangle
\vert \underline{x}\rangle_{cp}\longrightarrow \vert
\underline{y}\rangle \sum_{\underline{x}}c_{\underline{x}}\vert
\underline{x}\rangle_{cp}$ where $\vert \underline{y}\rangle$ is
the cleaned up state for the region $R$. The overall process is
reversible as it can be described by the transformation
$\sum_{\underline{x}}c_{\underline{x}}\vert \underline{x}\rangle
\vert \underline{0}\rangle_{cp} \longrightarrow \vert
\underline{y}\rangle \sum_{\underline{x}}c_{\underline{x}}\vert
\underline{x}\rangle_{cp}$. If the final state of the quantum
robot depends on the initial state of the systems in region $R$,
then correlations remain and the overall transformation
corresponding to carrying out the cleanup task is given by 
$\sum_{\underline{x}}c_{\underline{x}}\vert \underline{x}\rangle
\vert \underline{0}\rangle_{cp}\theta_{i} \longrightarrow \vert
\underline{y}\rangle \sum_{\underline{x}}c_{\underline{x}}\vert
\underline{x}\rangle_{cp}\theta_{\underline{x}}$.  Here
$\theta_{i}$ and $\theta_{\underline{x}}$ are the initial and
final states of the quantum robot.

Another type of task has the goal of carrying out measurements
or physical experiments on the environment.  Here the emphasis is
on the extraction or transfer of information from the environment
and not on a specified change of the state of the environment. 
An example of this type of task is "determine the distance
between particle (p) and the quantum robot (QR)".  If (p) and QR are in
respective position states $\vert x\rangle_{p}$ and $\vert
j\rangle_{QR}$ then carrying out this task corresponds to the
transformation $\vert j\rangle_{QR}\vert E_{x}\rangle \vert
i\rangle_{rec} \Longrightarrow \vert j\rangle_{QR}\vert
E_{x}^{\prime}\rangle \vert d(j,x)\rangle_{rec}$.  Here $\vert
i\rangle_{rec}$ and $\vert d(j,x)\rangle_{rec}$ denote the
initial and final states of the recording system where $d(j,x)$
denotes the distance between positions $j$ and $x$. The state
$\vert E_{x}\rangle  =\vert x\rangle_{p}\vert E\rangle_{\neq p}$
denotes the initial state of the environment with particle (p) at
position $\underline{x}$. Here $\vert E\rangle_{\neq p}$ is the
initial state of environment systems other than p and $\vert
E_{x}^{\prime}\rangle$ denotes the final state of all environment
systems including p after interaction of the quantum robot
at site $x$.  

Reversibility of this task requires that the final states $\vert
d(j,x)\rangle_{rec}\vert E^{\prime}_{x}\rangle$ be pairwise
orthogonal for different values of $j,\; x$.  This can be
achieved by requiring that the states $\vert d\rangle_{rec}$ are
pairwise orthogonal for different values of $d$ and are
orthogonal to $\vert i\rangle_{rec}$. Also for pairs of positions
$j,\;x$ and $j,\; x_{1}$ where $d(j,x) =d(j,x_{1}$ the states
$\vert E_{x}^{\prime}\rangle$ and $\vert
E^{\prime}_{x_{1}}\rangle$ should be orthogonal.  

In this paper the dynamics of each task is described as a 
sequence of alternating computation and action phases.  This is assumed to be
the case independent of the type or goal of the task.  The
purpose of each computation phase is to determine the action to
be taken by the quantum robot in the following action phase and possibly to
record local environmental information.  The
input to the computation, carried out by the on board quantum
computer, includes the local state of the environment and any
other pertinent information, such as the output of the previous
computation phase.  During a computation phase the quantum robot
does not move or change the state of the environment.  It does
change the state of an on board ancillary system, the output
system whose state determines the action taken following
completion of the computation.

During each action phase the state of the environment is changed
and the quantum robot can move. The state of the output system
(o) is not changed.  An action phase may consist of one or more
steps. During each step changes in the environment state are
limited to a neighborhod of the quantum robot.  Also an upper
bound is set on the distance the quantum robot can move during
each step. This is done to avoid jumps over arbitrary distances
by the quantum robot during a step.  

What happens during an action phase depends on the state of the
ouput system. It may also depend on the state of the neighborhood
environment of the quantum robot during any step. Examples of
actions that do not and do require observations are "move the
quantum robot one step in the $+x$ direction"  and "move the
quantum robot successive steps in the $+x$ direction as long as
no particles are encountered.  Do not move if a particle is
encountered." 

The description of tasks carried out by quantum robots requires
the use of completion or halting flags to determine when
individual action and computation phases are completed as well as
when the overall task is completed.  Such flags are necessary
because the unitarity of the time step operator requires that
system motion occurs somewhere even after the task is completed.

Note that there are many examples of tasks that never halt. 
Nonhalting of tasks can arise from several sources.  The task may
consist of a nonterminating sequence of computation and action
phases. Or either a computation phase or an action phase may
never halt.  An example of an action that is multistep, does not
halt, and requires local environment interactions at each step is
the above example in case the environment contains no particles
in the $+x$ direction from the quantum robot.

As noted the purpose of a computation phase is to determine the
action to be taken in the following phase.  It seems intuitively
reasonable to implement this determination by use of a quantum
computer on board the quantum robot.  However one can ask if quantum 
computers are really necessary
here.  Is it sufficient to limit consideration to special purpose
dedicated quantum robots that can carry out specific tasks or
groups of tasks in most any environment? This question is
emphasized by the fact that the model described in the next two
sections makes no explicit use of quantum computers.

A definite answer cannot be given at this point. However it is
likely that they are necessary. To support this one notes that it
is reasonable to require that for each task there exists a
physically reasonable $T$ such that each phase is implemented
efficiently. (That is the number of time steps is reasonable). 

The exact physical meaning of efficient implementation is not
clear at present. However the definition used in computer science
(computations dealing with numbers $\approx n$ are efficient if
the number of steps is polynomial in $\log n$) leads to the
following suggestion: Implementation of a phase (and a task) is
efficient if the number of steps needed to complete a phase is
polynomial in the number of relevant information bearing degrees
of freedom of the quantum robot. In particular it should not be
polynomial in the dimensionality of the Hilbert space of states
for the information bearing degrees of freedom as this
corresponds to being exponentially slow (polynomial in $n$).

Another requirement is based on the assumption that there should
exist a physically reasonable quantum robot that can carry out
almost any task efficiently in almost any environment. This is
equivalent to requiring the the existence, in principle at least,
a general purpose or universal quantum robot that can, with minor
modifications, carry out almost any task efficiently on almost
any environment. Minor modifications mean such things as use of
shielding for harsh environments, increase of the number of
information bearing degrees of freedom for complex tasks, etc.. 

It is suspected that such general purpose efficient quantum
robots require the presence of a universal quantum computer on board. The
type of quantum computer and number of relevant degrees of
freedom in the computer, as well as the need to carry out
efficient quantum computer algorithms such as those of Shor or
Grover \cite{Shor,Grover}, may depend on the task and environment
being considered.  However, these are all questions for the
future.

\section{A Model of Quantum Robots plus Environments}
\label{SMQRE}

Here a model of quantum robots interacting with environments is
described that illustrates the above material.  In the interests
of clarity and for purposes of illustration several simplifying
assumptions and limitations will be made.  First the model will
be limited to a description of information bearing degrees of
freedom only.  The relevance of this for the development of
quantum computers has been noted by Landauer \cite{Land1}. 

As noted a quantum robot (QR) contains a quantum
computer and ancillary systems on board.  The quantum computer can be
modelled as a cyclic network of quantum gates, a quantum Turing
machine, or by any other suitable method.  Since the material in
this section does not depend on any specific model, none will be
chosen here. Ancillary systems present are an output system (o),
and a control qubit (c). In addition a memory system may also be
present.  

Environments are considered to consist of arbitrary numbers and
types of particles on 1-,2-, or 3-D space lattices. Very simple
examples of environments consist of a 1-D lattice of qubits
(which is a quantum register) and a 1-D lattice containing just
one spinless particle.  Figure 1 shows a quantum robot in a 3-D
space lattice environment where the on board computer is a
quantum Turing machine.  Environment systems external to the
quantum robot are not shown.  The location of the quantum robot
in the lattice is shown by an arrow.  

Besides the assumption of discrete space and time, it is assumed that changes 
in the states of environment systems occur only as a result of interactions 
with the quantum robot. The states are stationary in the absence of this
interaction.  This restrictive assumption is made to avoid
dealing with complications in describing task dynamics for
environments of moving interacting systems. It is hoped to remove
this restrictive assumption in future work.

The assumed discreteness of time means that motion of the overall
system occurs in discrete time steps on a
space lattice.  Based on this a unitary step operator $T$ is
associated with each task where $T$ describes the task dynamics
for one time step. For each $n$ the system
dynamics for $n$ time steps in the forward (or backward) time
direction is given by $T^{n}$ (or $(T^{\dag})^{n}$).  

This association of $T$ with a finite time interval is similar to
the assumption made by Deutsch and others
\cite{Deutsch85,Deutsch89,BeVa} for quantum computers.
Alternatively $T$ can be associated with an infinitesimal time
interval.  In this case $T$ can be used to directly construct a
Hamiltonian according to \cite{Feynman}:
\begin{equation}
H=K(2-T -T^{\dag}) \label{ham}
\end{equation}
where $K$ is an arbitrary constant. In this model $T$ need not be
unitary or even normal $(TT^{\dag} \neq T^{\dag}T$ is possible). 
This model, which has been described in detail elsewhere for
quantum computers \cite{Benioff86,BenioffQBE}, will not be used
here.

The description of each task as a sequence of computation and
action phases is reflected in the separation of $T$ into
operators $T_{a}$ and $T_{c}$ describing single steps in action
phases and computation phases respectively for the quantum robot. 
That is 
\begin{equation}
T= T_{c} + T_{a} \label{Tsum}
\end{equation}

As noted before the goal of a computation phase ($T_{c}$ active) 
is to determine the action to be carried out in the following
action phase.  The states of (o) and the neighborhood environment are
input for the computation. The computation, which is in general
multistep, determines a new state of (o) as output. There is no change in 
the environment state or the location of the quantum robot.

The goal of the action phase ($T_{a}$ active) is to carry out the
action based on the state of (o). Actions include motion of the
quantum robot and local changes of the environment state.  They
may be single step or multistep and may or may not require local
observation of the environment. The states of (o) and the on
board quantum computer are not changed.

The function of the control qubit (c) is to regulate which type
of phase is active.  $T_{c}$ or $T_{a}$ is active if (c) is in
the respective  states $\vert 0\rangle$ or $\vert 1\rangle$.  The
last step, or iteration, of $T_{c}$ or $T_{a}$, of the
computation or action phase includes the respective change $\vert
0\rangle_{c}\rightarrow \vert 1\rangle_{c}$ or $\vert
1\rangle_{c}\rightarrow \vert 0\rangle_{c}$.

The conditions that $T_{c}$ and $T_{a}$ must satisfy can be
expressed in terms of properties of these operators relative to a
reference basis   ${\cal B} ={\cal B}^{qc}\otimes {\cal
B}^{anc}\otimes{\cal B}^{ext}$ for the quantum robot and
environment. Here ${\cal B}^{qc} = \{\vert b\rangle_{qc}\}$ is a
reference basis for the quantum computer. If the on board quantum
computer is a quantum Turing machine as in Figure 1, then  $\vert
b\rangle_{qc} =\vert m,k,\underline{t}\rangle$ where $\vert
m\rangle$ and $\vert k\rangle$ denote the respective internal
state and ${\cal L}_{2}$ location of the head $h_{2}$, and 
$\vert \underline{t}\rangle = \otimes_{j=1}^{N+1}\vert
\underline{t}_{j}\rangle$ is the state of the qubits on ${\cal
L}_{2}$ with $\underline{t}_{j} = 0,1$ for $N$ qubits and
$\underline{t}_{j} = 0,1,2$ for the marker qubit.   For the
ancillary systems ${\cal B}^{anc} = \{ \vert \ell_{1}\rangle_{o}
\vert i\rangle_{c}\}$ where $\{ \vert \ell_{1}\rangle_{o}\}$ is a
finite basis for the output system and $\{ \vert i\rangle_{c}\}$
with $i=0,1$ is a basis for the control qubit.  The external
basis ${\cal B}^{ext}$ for the environment systems and position of the 
quantum robot is given by $\{\vert \underline{x}\rangle_{QR}
\vert E\rangle \}$. The state $\vert \underline{x}\rangle_{QR} =
\vert x,y,z\rangle_{QR}$ gives the lattice site location of the
quantum robot, denoted by the arrow in Figure 1. The basis $\{
\vert E\rangle \}$ denotes a chosen basis for the environment of
quantum systems. 

The requirement that $T_{c}$ not change the environment state or
the QR location is given by 
\begin{equation}
T_{c} = \sum_{\underline{x},E}P^{e}_{\underline{x},E}
T_{c}P^{e}_{\underline{x},E}P^{c}_{0} \label{Tcext} 
\end{equation}
where $P^{e}_{\underline{x},E}=\vert \underline{x},E\rangle
\langle \underline{x},E\vert$ is the projection operator for the
QR at site $\underline{x}$ and the environment in state $\vert
E\rangle$.  This equation expresses the requirement that 
iteration of $T_{c}$ does not change the location of the quantum
robot or the state of the environment relative to the chosen
basis, (i.e. $T_{c}$  is diagonal in states $\vert
\underline{x},E\rangle$). 

This can also be expressed by the requirement $\langle
\underline{x^{\prime}} E^{\prime}\vert T_{c}\vert \underline{x}
E\rangle = T_{c}^{\underline{x}E} \langle
\underline{x^{\prime}}\vert \underline{x}\rangle \langle
E^{\prime}\vert E\rangle$ where $T_{c}^{\underline{x}E} = \langle
\underline{x} E\vert T_{c}\vert \underline{x} E\rangle$ is the
operator for the on board systems for the external state $\vert
\underline{x} E\rangle$.  The action of $T_{c}$ in the presence
of external states $\sum_{\underline{x},E}c_{\underline{x}E}
\vert \underline{x}E\rangle$ will in general introduce
entanglements between the external basis states and states of the quantum
computer. The presence of the projection operator $P^{c}_{0}$ for
the control qubit shows that $T_{c}$ is inactive if the control
qubit is in state $\vert 1\rangle$.

To express the requirement that the dependence of  $T_{c}$  on
the state of the environment is limited to the state of the
environment in a neighborhood of the quantum robot, one chooses
environment basis states that can be expressed as product of
states of systems inside and outside of neighborhoods.  For each
lattice position $\underline{x}$ of the quantum robot, let
$N(\underline{x})$ denote a neighborhood of $\underline{x}$. Then
the environmental basis can be chosen so that $\vert E\rangle =
\vert E\rangle_{N(\underline{x})}\rangle \vert E_{\neq
N(\underline{x})}\rangle$. Here $\vert
E\rangle_{N(\underline{x})}$ and $\vert E\rangle_{\neq
N(\underline{x})}$ are the states of environment systems in and
outside of $N(\underline{x})$. 

As a specific example let the environment consist of $n$
particles each with internal degrees of freedom. Then $\vert
E\rangle =\otimes_{j=1}^{n}\vert \underline{x_{j}} f_{j}\rangle$
where $\vert \underline{x_{j}}\rangle$ and $\vert f_{j}\rangle$
denote the lattice position state and the state of the internal
degrees of freedom of the jth particle. The state $\vert
E\rangle$ can also be written as 
\begin{equation}
\vert E\rangle = \otimes_{j=1}^{m}\vert
\underline{x_{\ell_{j}}}f_{\ell_{j}}\rangle \otimes_{h=1}^{n-m}\vert \underline{x_{k_{h}}}f_{k_{h}}\rangle = \vert
E\rangle_{N(\underline{x})}\rangle \vert E_{\neq
N(\underline{x})}\rangle \label{Eloc}
\end{equation}
where m of the n systems are inside $N(\underline{x})$ and the
rest are outside.

The requirement that $T_{c}$ depend on the environment only in
the neighborhood of the quantum robot can be be expressed by the
condition
\begin{equation}
\langle \underline{x},E\vert T_{c}\vert \underline{x},E\rangle
=\langle E_{N(\underline{x})}\vert T_{c}\vert
E_{N(\underline{x})}\rangle \label{Tcextloc}
\end{equation} 
for the quantum robot operator.  Here Eq. \ref{Eloc} and $\langle
E_{\neq N(\underline{x})}\vert E_{\neq N(\underline{x})}\rangle
=1$ have been used.  This condition is equivalent to requiring that $T_{c}$
is the identity on the space of states of environment systems outside of
$N(\underline{x})$.

The dependence of $\langle E_{N(\underline{x})} \vert T_{c}\vert
E_{N(\underline{x})}\rangle$ on the neighborhood environmental
states can be very complex as it can depend on all the $m$
variables $f_{\ell_{1}}, \cdots ,f_{\ell_{m}}$ of Eq. \ref{Eloc}
as well as on which of the $n$ systems are inside
$N(\underline{x})$. If the $n$ environmental systems are all
fermions or bosons, then the complexity is reduced because of
symmetry restrictions on the environmental states.    For example
for fermions if the neighborhood $N(\underline{x})$ is just the
point $\underline{x}$ and $f$ can assume $M$ values, then there
are $2^{M}$ distinct environmental states $\vert
E_{\underline{x}}\rangle$ (provided  $n>M$). By Eq.
\ref{Tcextloc} $\langle E_{N(\underline{x})} \vert T_{c}\vert
E_{N(\underline{x})}\rangle$ can be different for each of these
states.  If the particles are bosons then there are even more
distinct local environment states possible as an arbitrary number
of systems in the same internal state can be present at the QR
location and $T_{c}$ may depend on the number of systems present.

Note that the above description includes a distinct value for 
$\langle E_{N(\underline{x})} \vert T_{c}\vert
E_{N(\underline{x})}\rangle$ in the case that no systems are in
$N(\underline{x})$. This describes the computation phase operator
if the neighborhood environment is empty.  Tasks that include
search operations in an environment to find systems make use
of this phase, especially if the environment is sparsely
populated.

Much of the above discussion also applies to the action phase
operator $T_{a}$.  This operator depends on but does not change
the states of (o) (and a memory system if present) relative to
some basis.  This condition can be expressed by an equation
similar to Eq. \ref{Tcext}:
\begin{equation}
T_{a} = \sum^{\prime}_{\underline{x^{\prime}},\underline{x}}
\sum_{\ell_{1}} P^{qr}_{\underline{x^{\prime}}}
P^{o}_{\ell_{1}}T_{a}P^{o}_{\ell_{1}} P^{qr}_{\underline{x}}
P^{c}_{1}.
 \label{Taext}
\end{equation}
where $P^{o}_{\ell_{1}}$ is the projection operator for (o) in
state $\vert \ell_{1}\rangle$, $P^{qr}_{\underline{x}}$ is the
projection operator for the quantum robot at lattice location
$\underline{x}$, and $P^{c}_{1}$ is the projection operator for 
(c) in state $\vert 1\rangle$. These conditions show that $T_{a}$
is diagonal in the states $\vert \ell_{1}\rangle_{o}$ and is
inactive when (c) is in state $\vert 0\rangle$.

The limitation on the sum over quantum robot positions, shown by 
the prime on  $\sum^{\prime}_{\underline{x^{\prime}},
\underline{x}}$ expresses the restriction to one site motion in
any direction for the quantum robot during one step. That is, if
$\underline{x^{\prime}} = (x^{\prime},y^{\prime},z^{\prime})$ and
$\underline{x} = (x,y,z)$ then $(x^{\prime}=x\pm 1,
y^{\prime}=y,z^{\prime}=z)$ or $(x^{\prime}=x,y^{\prime}=y\pm
1,z^{\prime}=z)$ or $(x^{\prime}=x,y^{\prime}=y,z^{\prime}=z\pm
1)$ or $\underline{x{\prime}} = \underline{x}$ are possible along
with linear superpositions of these seven alternatives.  

$T_{a}$ is independent of both the states of the on board
quantum computer (qc) and the states of environment systems
distant from the quantum robot.  This means that $T_{a}$ is the identity 
on the space spanned by all the states in ${\cal B}^{qc}$. In addition,
\begin{equation}
\langle \underline{x^{\prime}} E^{\prime}\vert T_{a} \vert
\underline{x} E\rangle = \langle E_{\neq
N(\underline{x^{\prime}},\underline{x})}^{\prime}\vert E_{\neq
N(\underline{x^{\prime}},\underline{x})}\rangle \langle
\underline{x^{\prime}} E_{N(\underline{x^{\prime}},
\underline{x})}^{\prime}\vert T_{a} \vert \underline{x}
E_{N(\underline{x^{\prime}},\underline{x})}\rangle \label{Taext1}
\end{equation}
for all $ \underline{x^{\prime}},\; \underline{x}$ such that
$\vert \underline{x^{\prime}} - \underline{x}\vert \leq 1$. The
states $\vert E_{N(\underline{x^{\prime}},
\underline{x})}\rangle$ and $\vert E_{\neq
N(\underline{x^{\prime}}, \underline{x})}\rangle$ describe the
respective environments inside and outside the combined
nieghborhoods of $\underline{x^{\prime}}$ and $\underline{x}$. 
The definition of these states is similar to that given earlier,
Eq. \ref{Eloc}, for $\vert E_{N(\underline{x})}\rangle$ and
$\vert E_{\neq N(\underline{x})} \rangle$. Also $\vert E\rangle =
\vert E_{N(\underline{x^{\prime}}, \underline{x})}\rangle \vert
E_{\neq N(\underline{x^{\prime}}, \underline{x})}\rangle$ has
been used. The matrix element $\langle E_{\neq
N(\underline{x^{\prime}}, \underline{x})}^{\prime}\vert E_{\neq
N(\underline{x^{\prime}},\underline{x})}\rangle =1$ if and only
if $\vert E^{\prime}\rangle = \vert E\rangle$ st sites outside
$N(\underline{x^{\prime}},\underline{x})$.  Otherwise it equals
$0$.  

The righthand matrix element of Eq. \ref{Taext1} expresses the
limitation that one action phase step can change the environment
at most in the neighborhoods of the initial and final locations
of the quantum robot. As noted earlier, motion of the quantum
robot is limited to at most one lattice site in any direction. 
If desired these limitations can be relaxed by suitable
modifications of Eqs. \ref{Taext} and \ref{Taext1}. 

Several additional aspects of the properties of $T_{a}$ and
$T_{c}$ need to be noted. One is that to avoid complications, the
need for history recording has not been discussed. Both the
computation and action phases may need to record some history.
For example when $T_{c}$ is active, the change $\vert
\ell\rangle_{o} \longrightarrow \vert \ell^{\prime}\rangle_{o}$
requires history recording if the change is not reversible. 
Where records are stored (on board the quantum computer or in the
environment) depends on the model. Also the task carried out by
the quantum robot may not be reversible unless the components of the initial
state of the relevant regions of the environment is copied or recovered. 

Initial and final states for the starting and completion of 
tasks may be needed. For example, at the outset, the output and
control systems might be  in the state $ \vert
\ell_{i}\rangle_{o}\vert 0\rangle_{c}$ and the environment would
be in some suitable initial state.   The process begins with the
on board quantum computer active.

Completion of a task could be described by designating one or
more states $\vert \ell_{f}\rangle$ as final output states and
arranging matters so that motion of some type occurs that does
not destroy the final task state. This ballast motion can occur
on board the quantum computer or consist of motion of the quantum
robot or some other system along a path in the environment
without changing the environmental state, or it can be a combination of
both. If the ballast motion occurs on board the quantum computer
and it is described by states in a finite dimensional Hilbert
space, the stability of the final task state lasts for a finite
time only before the task is undone.

The conditions given above for $T_{c}$ and $T_{a}$ are
sufficiently general to allow for branching tasks with states
describing entangled activities.  For example during a
computation phase $T_{c}$ can take an (o) state $\vert \ell
\rangle$ into a linear superposition
$\sum_{\ell_{1}}c_{\ell_{1}}\vert \ell_{1}\rangle$. Similarly the
action of $T_{a}$ can take an environment and QR position state
$\vert \underline{x},E\rangle$ into a linear superposition
$\sum_{\underline{x^{\prime}}E^{\prime}}c_{\underline{x^{\prime}}E^{\prime}}\vert \underline{x^{\prime}}E^{\prime}\rangle$. In
this case the sum is limited to values of $\underline{x^{\prime}}
E^{\prime}$ that satisfy Eq. \ref{Taext} and \ref{Taext1}. Additional 
branching is possible if the action of $T_{c}$ or
$T_{a}$ takes control qubit states into linear sums of $\vert
0\rangle$ and $\vert 1\rangle$.  This allows for entanglements of
action and computation phases.

\section{Sum Over Phase Paths}
\label{SOP}

Another quite illuminating way to study the time development of
the model implementation of a given task is by use of the sum
over paths method \cite{FeHi}.  If $\Psi(0)$ and $\Psi(n)$ are
the respective overall system initial state and state after n
time steps, then $\Psi(n) =T^{n}\Psi(0)$. In particular the
amplitude that one ends up in state $\vert
b,\ell,i,\underline{x},E\rangle$ is given by 
\begin{equation}
\langle \Psi(n)\rangle =
\sum_{b_{1},\ell_{1},i_{1},\underline{x_{1}},E_{1}}\langle 
b,\ell,i,\underline{x},E \vert T^{n} \vert
b_{1},\ell_{1},i_{1},\underline{x_{1}},E_{1}\rangle \langle
b_{1},\ell_{1},i_{1},\underline{x_{1}},E_{1}\vert \Psi(0)\rangle.
\end{equation}

As is well known, the matrix element $\langle 
b,\ell,i,\underline{x},E \vert T^{n} \vert
b_{1},\ell_{1},i_{1},\underline{x_{1}},E_{1}\rangle$, that gives
the amplitude for evolving from state $\vert
b_{1},\ell_{1},i_{1},\underline{x_{1}},E_{1}\rangle$ to state
$\vert b,\ell,i,\underline{x},E\rangle$ in $n$ steps plays a very
important role in a description of the time development of the
system.  To simplify notation let the state $\vert w,i\rangle$
denote the state $\vert b,\ell,i,\underline{x},E\rangle$.

Expansion in a complete set of states between each $T$ factor
gives
\begin{equation}
\langle w,i\vert T^{n}\vert w_{1},i_{1}\rangle =
\sum_{w_{2},i_{2},\cdots ,w_{n},i_{n}} \langle w,i\vert T\vert
w_{n},i_{n}\rangle \langle w_{n},i_{n}\vert T\vert w_{n-1},i_{n-1}\rangle, \cdots ,\langle w_{2},i_{2}\vert T\vert w_{1},i_{1}
\rangle. \label{intsum}
\end{equation}
This can also be written as a sum over paths of states $\{\vert
w,i\rangle \}$ of length $n+1$ whose initial and final elements
are $\vert w_{1},i_{1}\rangle$ and $\vert w,i\rangle$
\cite{FeHi}:
\begin{equation}
\langle w,i\vert T^{n}\vert w_{1},i_{1}\rangle
=\sum_{\stackrel{\scriptstyle{paths \; p\; \; of}}
{length \; n+1}} \langle p_{n+1}\vert
T\vert p_{n}\rangle,\cdots , \langle p_{2}\vert T\vert
p_{1}\rangle \langle p_{n+1}\vert w,i\rangle \langle p_{1}\vert
w_{1},i_{1}\rangle. \label{pathsum}
\end{equation}

In this paper tasks are defined as sequences of alternating
computation and action phases. To make this feature explicit it
is necessary to separate out sums over control qubit states. 
Since 
\begin{equation}
T^{n} = (P_{0}+P_{1})T(P_{0}+P_{1})T(P_{0}+P_{1}),\cdots
,(P_{0}+P_{1})T(P_{0}+P_{1}) \label{Tsumcqu}
\end{equation}
where $P_{i}$ is the (c) qubit projection operator for state
$\vert i\rangle_{c}$, one can use the fact that, by Eqs.
\ref{Tsum}, \ref{Tcext}, and \ref{Taext}, $T_{a} =TP_{1}$ and
$T_{c} = TP_{0}$ to write 
\begin{equation}
T^{n}= \sum_{v_{1} =a,c}\sum_{t=1}^{n}\sum_{h_{1},h_{2},\cdots
,h_{t} =1}^{\delta (\sum ,n)}(P_{0}+P_{1})(T_{v_{t}})^{h_{t}}
(T_{v_{t-1}})^{h_{t-1}},\cdots
,(T_{v_{2}})^{h_{2}}(T_{v_{1}})^{h_{1}}. \label{Tacsum}
\end{equation}
Here $v_{j+1} = a$ (or $c$) if $v_{j}=c$ (or $a$). The upper
limit $\delta (\sum ,n)$ on the $t$ fold sum over $h_{1},\;
h_{2},\cdots ,h_{t}$ means that the sum is limited to values that
satisfy $h_{1}+h_{2} +,\cdots ,+h_{t} =n$.

This equation shows explicitly the expansion of $T^{n}$ as a sum
of alternating completed computation and action phase operators.
The term for each value of $t$ and each value of $h_{1}, \cdots
,h_{t}$ corresponds to a sequence of $t$ alternating computation
and action phases consisting of $h_{1}, h_{2}, \cdots ,h_{t}$
steps. The operators for each phase are time ordered in that
$(T_{v_{j+1}})^{h_{j+1}}$ occurs after $(T_{v_{j}})^{h_{j}}$. 
Note that $T_{a}$ and $T_{c}$ do not commute. If $v_{1}=c$ then
the sequence begins with $T_{c}$.  It ends with $T_{a}$ (or
$T_{c}$) if $t$ is even (or odd).  For example if $v_{1}=c$ and
$t$ is even the terms in Eq. \ref{Tacsum} have the form
\begin{displaymath}
T_{a}^{h_{t}}T_{c}^{h_{t-1}},\cdots ,T_{a}^{h_{2}}T_{c}^{h_{1}}.
\end{displaymath}
If $v_{1} =a$ then $a$ and $c$ are interchanged in the
alternation. The terminal factor $P_{0}+P_{1}$ allows for
termination or extension of the phase associated with
$T_{v_{t}}$.  Note that the sums include terms for just one
action or computation phase with $n$ steps up to maximal
alternation of $n$ computation and action phases each with just
one term.

It is useful to expand the amplitude $\langle w,i\vert T^{n}\vert
w_{1},0\rangle$ as a sum over states at the beginning and end of
each phase.  This can be done using Eq. \ref{Tacsum} to obtain
\begin{equation}
\langle w,i\vert T^{n}\vert w_{1},0\rangle =
\sum_{t=1}^{n}\sum_{w_{2},\cdots ,w_{t}} \sum_{h_{1},h_{2},\cdots
,h_{t} =1}^{\delta(\sum ,n)}\langle w,i\vert
(T_{v_{t}})^{h_{t}}\vert w_{t}\rangle ,\cdots ,\langle w_{3}\vert
(T_{a})^{h_{2}}\vert w_{2}\rangle \langle w_{2}\vert
(T_{c})^{h_{1}}\vert w_{1}\rangle \label{Tintsum}
\end{equation}
where, as before, $\vert w\rangle$ denotes $\vert
b,\ell,\underline{x},E\rangle$.

Each term in this large sum gives the amplitude for finding $t$
alternating phases in the first $n$ steps such that each of the
$t$ phases begins with a specified input state and ends after a
specified number of steps in a specified output state.  The sums
over $h_{1} \cdots ,h_{t}$ have been commuted past the state sums
over $w_{2},\cdots ,w_{t}$ as it is merely a rearranging of
terms. 

 As was done for Eq. \ref{pathsum} the sum over $w_{2}, \cdots
,w_{t}$ can be replaced by a sum over length $t+1$ paths of
states where the initial and final states of each path are $\vert
w_{1}\rangle$ and $\vert w\rangle$. In particular one has
\begin{equation}
\langle w,i\vert T^{n}\vert w_{1},0\rangle =
\sum_{t=1}^{n}\sum_{\stackrel{\scriptstyle{paths \; p\;
\; of }} {length \; t+1}}
\sum_{h_{1},h_{2},\cdots ,h_{t} =1}^{\delta(\sum ,n)}\langle
p_{t+1},i\vert (T_{v_{t}})^{h_{t}}\vert p_{t}\rangle ,\cdots
,\langle p_{3}\vert (T_{a})^{h_{2}}\vert p_{2}\rangle \langle
p_{2}\vert (T_{c})^{h_{1}}\vert p_{1}\rangle \langle w\vert
p_{t+1}\rangle \langle p_{1}\vert w_{1}\rangle 
\label{Tphpathsum}
\end{equation}
where $\vert p_{j}\rangle = \vert w_{j}\rangle = \vert
b_{j},\ell_{j}, \underline{x_{j}},E_{j}\rangle$ denotes the $j$th
state in path $p$.

This result is quite useful in that it expresses the amplitude
$\langle w,i\vert T^{n}\vert w_{1} 0\rangle$ as a sum over phase
paths containing $t$ phases where $1\leq t\leq n$.  Included are
sums over different numbers of steps for each phase subject to
the condition that the total number of steps is $n$. The sums
over state paths describing motion within each phase are
suppressed.

The conditions on $T_{a}$ and $T_{c}$, expressed in Eqs.
\ref{Taext} and \ref{Tcext}, have the consequence that many of
the path amplitudes in Eq. \ref{pathsum} and phase path
amplitudes in Eq. \ref{Tphpathsum} do not contribute. Because of
this the path sums and phase paths sums can be restricted to only
those paths or phase paths  that satisfy the conditions on
$T_{a}$ and $T_{c}$.

Additional restrictions on the phase path sum derive from the
fact that for a given task $T$ is supposed to implement the task. 
For example suppose the task is such that a decision tree can be
associated with the task where the tree shows the temporal
ordering, alternatives, and desired outcomes of task steps based
on outcomes of prior steps. The decision tree limits the sum over
phase paths to those paths that are consistent with the paths in
the tree (and with the requirement that a task is a sequence of
computation and action phases).  Other paths have $0$ amplitudes
(at least if $T$ is error free).

For many simple tasks any $T$ that implements the task is such 
that just one phase path has nonzero amplitude.  In this case Eq.
\ref{Tphpathsum} becomes
\begin{equation}
\langle w,i\vert T^{n}\vert w_{1},0\rangle =
\sum_{t=1}^{n}\sum_{h_{1},h_{2},\cdots ,h_{t} =1}^{\delta(\sum
,n)}\langle \tilde{p}_{t+1},i\vert (T_{v_{t}})^{h_{t}}\vert
\tilde{p}_{t}\rangle ,\cdots ,\langle \tilde{p}_{3}\vert
(T_{a})^{h_{2}}\vert \tilde{p}_{2}\rangle \langle
\tilde{p}_{2}\vert (T_{c})^{h_{1}}\vert \tilde{p}_{1}\rangle
\langle w\vert \tilde{p}_{t+1}\rangle \langle \tilde{p}_{1}\vert
w_{1}\rangle  \label{onepath}
\end{equation}
where $\tilde{p}$ denotes the contributing path.  The $t$ sum is
over initial segments of length $t$ of the path $\tilde{p}$. The
sums over $h_{1},\cdots ,h_{t}$ express the fact that in general
there is neither a definite completion time nor a definite 
duration time for each phase.  The dependence of the amplitude
factors $\langle \tilde{p}_{j+1}\vert (T_{v_{j}})^{h_{j}}\vert
\tilde{p}_{j}\rangle$ on $h_{j}$ depends on $T$ and the phase
path states.

\section{A Very Simple Example}
\label{AVSE}

Here the very simple example described in Section \ref{QR} of
determination of the distance between the quantum robot and a
system will be will be considered to illustrate some aspects of
the models discussed above.  The environment is extremely simple
in that it consists of one spinless particle (p) on a 1-D space
lattice.  The task is carried out by the quantum robot moving to
the right on the lattice and counting the number of steps or
lattice sites as it moves. If the particle is located the number
of steps is recorded as the distance and the quantum robot
returns to its initial position and the task is completed. 

As noted earlier the overall quantum robot plus environment state
transformation resulting from carrying out the task can be
represented as
$\vert j\rangle_{QR}\theta(i)\vert x\rangle_{p}\longrightarrow
\vert j\rangle_{QR}\theta(x-j)\vert x\rangle_{p}$ provided the
particle is found.  Here $\vert j\rangle_{QR}\vert x\rangle_{p}$
denote the respective initial lattice positions of the quantum robot and
the particle, and $\theta(i)$ denotes the initial state of
internal degrees of freedom of the quantum robot. The state
$\theta(x-j)$ is the final internal state of the quantum robot
with the distance $x-j$ recorded in the memory.

If the initial state is a linear superposition of QR and (p)
position states the overall task transformation is given by
\begin{equation}
\Psi_{i} =\sum_{j,x}c_{j,x}\vert j\rangle_{QR}\vert
x\rangle_{p}\theta(i)\Rightarrow \Rightarrow
\sum_{j,x}^{\prime}c_{j,x}\vert j\rangle_{QR}\vert x\rangle_{p}
\theta(x-j) + \psi_{nf} \label{tasktr}
\end{equation}
The prime on the sum means that it is limited to values of $x-j$
such that $0 \leq x-j \leq 2^{N}-1$.  For these values the
quantum robot will find the particle. What happens if $x-j$ is
outside this range (the particle is not found) depends on model
assumptions. The state $\psi_{nf}$ represents the the task
transformation if the particle is not found. The states
$\theta(d)$ are pairwise orthogonal for different values of $d$
and are orthogonal to the initial state $\theta(i)$.

The description of the task in Eq. \ref{tasktr} and the
requirement of pairwise orthogonality of the states $\theta(d)$
ensure that the task is reversible except for the indeterminacy
resulting from which side (right or left) of the quantum robot
the particle is located. This is removed by specifying the
direction in which the search takes place.

For carrying out this task the on board quantum computer will be
considered to be a quantum Turing machine.  The quantum register
for the computer is taken to be a finite closed lattice  ${\cal
L}_{2}$ containing $N+2$ qubits: $N$ qubits are used for numbers
$0,1,\cdots ,2^{N}-1$, one qubit, which is ternary, is a marker,
and the remaining qubit adjacent to the marker denotes the sign
of the number ($\vert 1\rangle \sim +,\; \vert 0\rangle \sim -$).
This lattice will be used as a short term memory to keep a
running count of the number of sites the quantum robot moves at
each step.  

Another ancillary memory sytem (m) is added to the quantum robot.
This system consists of another $N+2$ qubit lattice ${\cal
L}_{3}$ like ${\cal L}_{2}$.  It is used to record permanently
the distance $x-j$ between the initial location of the QR and (p)
and corresponds to $\theta(x-j)$ in Eq. \ref{tasktr}. Figure 2
shows the setup on a 1-D lattice environment. 

There are three types of actions carried out in action phases for
this task: move to the right (mr), move to the left (ml), and do
nothing (dn). There are also two variants of the motion phases
used, move 1 lattice site and move without stopping. 
Corresponding to these, the output system (o) has five internal
states $\vert mr1\rangle_{o},\; \vert mr> \rangle_{o},\;
\vert ml1\rangle_{o},\;  \vert ml>\rangle_{o},\; \vert
dn\rangle_{o}$. The move right and left action phases for one
site  carry out the transformations $\vert j\rangle_{QR}\vert
x\rangle_{p}\rightarrow \vert j+1\rangle_{QR}\vert x\rangle_{p}$
and $ \vert j\rangle_{QR}\vert j\rangle_{p}\rightarrow \vert j-1\rangle_{QR}\vert x\rangle_{p}$ and stop. Do nothing means the
action phase makes no change in the QR and (p) position states. 
All these actions, and the nonstopping motions of the quantum
robot, do not involve environment observations.

The task begins with the number $+0$ on both on board lattices
and (o) in state $\vert dn\rangle_{o}$ and the computation phase
active.  If the particle (p) is at the QR location, the
computation subtracts $1$ from $0$ on the running memory lattice
${\cal L}_{2}$ and does not change in the state of (o). If (p) is
not at the location of QR, the computation phase adds $1$ to the
running memory and changes the (o) state to $\vert
mr1\rangle_{o}$.  In this case the subsequent action phase shifts
the QR $1$ site to the right and the computation phase becomes
active again.  

This stepwise process of adding $1$ to the number on the running
memory with no change in the (o) state $\vert mr1\rangle_{o}$ in
the computation phase, and one site QR motion in the action phase
continues until (p) is located. At this point the computation
phase copies the number from running memory to the permanent
memory ${\cal L}_{3}$, subtracts $1$ from the running memory, and
changes the (o) state to $\vert ml1\rangle_{o}$.  The next action
phase consists of moving the quantum robot back one lattice site.

This process continues until the number $0$ appears on the
running memory as part of the input to a computation phase. This
computation subtracts $1$ from the running memory and changes the
state of (o) to $\vert dn\rangle_{o}$.  At this point the task is
completed and the ballast phase begins.  Here ballast phase
motion consists of repeated subtraction of $1$ from the running
memory with intervening do nothing action phases.  The ballast
phase ends when the number $-(2^{N}-1)$ is in the running memory.

The task dynamics described above is shown schematically in
Figure 3 as a decision tree. The round circles $mr1,\;
mr>,\; ml1,\; ml>$, and $dn$ denote action phases. The
square boxes between successive action phases, denote memory
system states (d = running memory and st = permanent memory), and
questions with answers based on local environmental states.  The
collection of boxes and arrows between successive actions shows
what is done during each computation phase. The left hand column
shows the dynamics during the search part of the task.  The
central column, with horizontal arrows only, shows changes made
in memory states when the particle (p) is found, and the
righthand column shows the dynamics during the return part of the
task. The righthand row at the top shows progress during the
ballast part of the task.

In Figure 3 both the first column (the search phase) and the top
row (the ballast phase) end with the nonterminating action phases
$mr>$ and $ml>$ respectively. The $mr>$ action
which moves the quantum robot continually to the right, occurs in
case (p) is not located during the search phase.  This happens if 
(p) is either to the left or at least $2^{N}$ sites distant to
the right from the quantum robot.  The $ml>$ action phase,
which moves the quantum robot continually to the left, occurs at
the end of the ballast phase when the running memory is full of
$1$s.

Any $T$ that satisfies Eqs. \ref{Tsum}, \ref{Tcext}, and
\ref{Taext} and the conditions of the tree is such that iteration
of $T$ on a suitable initial state implements the task.  To see
this let the initial state $\Psi (0)$ be given by 
\begin{equation}
\Psi (0) = \vert j\rangle_{QR}\vert x\rangle_{p}\vert \underline{
0}\rangle_{{\cal L}_{3}}\vert \underline{0}\rangle_{{\cal L}_{2}}\vert
00\rangle_{h_{2}} \vert dn\rangle_{o}\vert 0\rangle_{c} = \vert
j\rangle_{QR}\vert x\rangle_{p}\vert \underline{0}\underline{0}00\rangle_{qtm} 
\vert dn\rangle_{o}\vert 0\rangle_{c}. \label{psiinit}
\end{equation}
This state expresses the initial conditions of the quantum Turing
machine given by the upper left hand corner or the decision tree
with the running memory ${\cal L}_{2}$ and permanent memory
${\cal L}_{3}$ lattices in states $\vert \underline{0}\rangle$,
the head $h_{2}$ in internal state $\vert 0\rangle$ and at the
location of the marker qubit on ${\cal L}_{2}$. The ouput and
control systems are in state $\vert dn,0\rangle$ and the
positions of the quantum robot and particle (p) are given by
$\vert j,x\rangle$.

The requirement that $T$ implement the task or decision tree of
Figure 3 means that for the initial state of Eq.\ref{psiinit}
just one term in the phase path sum of Eq. \ref{Tphpathsum} is
nonzero and that term corresponds to the specific path in the
decision tree that is followed for the initial state of Eq.
\ref{psiinit}.  This gives
\begin{equation}
 T^{n}\Psi (0) = \sum_{t=1}^{n}\sum_{h_{1},h_{2},\cdots ,h_{t}
=1}^{\delta(\sum ,n)} (T_{v_{t}})^{h_{t}}\vert
\tilde{p}_{t}\rangle \langle \tilde{p}_{t}\vert (T_{v_{t-1}})^{h_{t-1}}\vert \tilde{p}_{t-1}\rangle ,\cdots ,\langle
\tilde{p}_{3}\vert (T_{a})^{h_{2}}\vert \tilde{p}_{2}\rangle
\langle \tilde{p}_{2}\vert (T_{c})^{h_{1}}\vert
\tilde{p}_{1}\rangle \langle \tilde{p}_{1}\vert \Psi (0)\rangle
\label{Texpathsum}
\end{equation}
where as before $v_{t} =a,c$. This limitation to one path applies
only to paths of length $t$ without the terminal $t+1$st state as
the last factor $(T_{v_{t}})^{h_{t}}\vert \tilde{p}_{t}\rangle$
may not correspond to a completed phase.

The states in the path $\tilde{p}$ can be written down by
inspection of the decision tree and the initial state. If $x=j+2$ 
one has $\vert \tilde{p}_{1}\rangle =\vert j,j+2\rangle \vert
\underline{0},\underline{0},0,0\rangle_{qtm}\vert dn\rangle_{o}\vert 
0\rangle_{c},\;\vert \tilde{p}_{2}\rangle = \vert j,j+2\rangle \vert
\underline{0},\underline{1},0,0\rangle_{qtm}\vert mr1\rangle_{o}\vert 
1\rangle_{c},\;\vert \tilde{p}_{3}\rangle = \vert j+1,j+2\rangle \vert
\underline{0},\underline{1},0,0\rangle_{qtm}\vert mr1\rangle_{o}\vert 
0\rangle_{c},\;\vert \tilde{p}_{4}\rangle = \vert j+1,j+2\rangle \vert
\underline{0},\underline{2},0,0\rangle_{qtm}\vert mr1\rangle_{o}\vert 
1\rangle_{c},\;\vert \tilde{p}_{5}\rangle = \vert j+2,j+2\rangle \vert
\underline{0},\underline{2},0,0\rangle_{qtm}\vert mr1\rangle_{o}\vert 
0\rangle_{c},\;\vert \tilde{p}_{6}\rangle = \vert j+2,j+2\rangle \vert
\underline{2},\underline{1},0,0\rangle_{qtm}\vert ml1\rangle_{o}\vert 
1\rangle_{c}$. This
last state shows changes made by the computation phase at the end
of the search when the quantum robot is at the location of (p). 
The distance $2$ has been copied to the permanent record, $1$
subtracted from the running memory and the state of (o) changed
to $\vert ml1\rangle_{o}$.  Additional phase path states can be
found from the decision tree.

\section{Discussion}
\label{D}

Some aspects of the sum over paths need discussion. First it
should be noted that the decision tree of Figure 3 refers to a
quantum mechanical process, not a classical process. One
consequence is that there are no definite completion times or
durations for any of the phases corresponding to steps in the
tree.  This is the case even if the initial state has the quantum
robot and particle (p) in definite positions as in Eq.
\ref{psiinit} and just one path contributes. However, the decision tree does 
show the time ordering of the steps.

The lack of definite completion and duration times follows from
the fact that for any phase, such as the $j$th, on any path the amplitude
factor  $\langle p_{j+1}\vert (T_{v_{j}})^{h_{j}}\vert
p_{j}\rangle$ can be nonzero for many different values of
$h_{j}$.  The dependence of this factor on $h_{j}$, gives the
uncertainty in the duration time of the $j$th phase on path $p$.  If the
dependence is narrow and strongly peaked around some values the
uncertainty is small. If the dependence is broad and spread over
many values of $h_{j}$ the uncertainty is large.

Another point is that if the sum over phase paths contains more
than one path, the decision tree applies separately to each path. 
For the example studied this occurs if the initial state is a
linear superposition of states of the form given by Eq.
\ref{psiinit}.  This can also occur in case branchings occur in a
phase. For example suppose $T$ is such that the $m$th phase
branches with $T^{h_{m}}\vert p_{m}\rangle = \alpha \vert p_{m+1}
\rangle + \beta \vert p^{\prime}_{m+1}\rangle$ where $\alpha\neq
0 \neq \beta$. Here $p^{\prime}$ is another path that has the
first $m$ elements in common with $p$ and differs at the $m+1$st. 
 In this case and in more general sums over paths the peak values
and spreads in duration amplitudes for the phases can be quite
different in each of the paths. 

This branching may be an essential part of the task or it may be
due to errors in the construction of $T$. For instance, in the
example task, suppose that each time the action phase $mr1$ is
active it moves the state $\vert j\rangle_{qr}$ to a linear
superposition of $\vert j+1\rangle_{qr}$ and $\vert
j+2\rangle_{qr}$.  This could occur because of errors or
approximations in construction of $T$.  In this case the
expression of $T^{n} \Psi(0)$ as a sum over phase paths will
contain many paths instead of just one as in Eq.
\ref{Texpathsum}. The structure of the sum over phase paths is a
branching tree with binary branchings occurring whenever $mr1$ is
active. 

In this case the decision tree of Figure 3 applies to each phase
path separately  as it shows the sequence of actions and
computations that occur in each path. Of course errors will be
made in carrying out the task because for many paths the distance
recorded in the permanent memory (if one is recorded) will not
correspond to the actual distance between the quantum robot and
the particle (p). The total error amplitude consists of the sum
over all phase paths containing at least one $mr1$ phase
transformation of the form $\vert j\rangle_{qr}\rightarrow \vert
j+2\rangle_{qr}$.

As seen in Eq. \ref{pathsum} the amplitude for each phase path is
a product of single phase amplitudes. The structure of these
individual amplitudes is of interest in that they also can be
written as sums over variable length paths within each phase. For
example consider the $dn$ (do nothing) action phase in Figure 3.
One has
\begin{equation}
\sum_{m}\langle j,x,b,dn,0\vert T_{a}^{m}\vert j,x,b,dn,1\rangle
= \sum_{m}\sum_{\stackrel{ \scriptstyle{paths \; q\; \;
of}} {length \; m+1}}\langle
q_{m+1}\vert T_{a}\vert q_{m}\rangle \cdots \langle q_{2}\vert
T_{a}\vert q_{1}\rangle \langle j,x,b,dn,0\vert q_{m+1}\rangle
\langle q_{1}\vert j,x,b,dn,1\rangle \label{dnpthsum}
\end{equation}
where $T_{a} = TP^{c}_{1}$ has been used.  The state $\vert
j,x,b,dn,1\rangle$ refers to the quantum robot and particle (p)
at positions $j,x$, the quantum computer including permanent
memory in state $\vert b\rangle$, and the output and control
systems in states $\vert dn,1\rangle_{o,c}$. The (c) qubit output
state $\vert 0\rangle$ shows that these are amplitudes for
completed action phases.

This shows that the individual "do nothing" action phase
amplitudes are sums over paths of variable length with the
requirement that, except for the control qubit, the initial and
final path states are the same.  They correspond to  doing
nothing'.  On the other hand, no such requirement is needed for
the intermediate path states. The state $\vert q_{k}\rangle$ for
$1<k<m+1$ can be any basis state $\vert j^{\prime},x^{\prime},
b^{\prime},\ell^{\prime}, 1\rangle$.  Paths can wander anywhere
provided they begin and end in states corresponding to doing
nothing and satisfy the conditions on $T_{a}$ in Eq. \ref{Taext}.

This applies to completed computation and action phase amplitudes
in general. As discussed earlier, completed phase amplitudes must
begin and end with states describing changes appropriate to the
phase being considered.  Each phase path amplitude factor
$\langle p_{j+1}\vert (T_{v_{j}})^{h_{j}}\vert p_{j}\rangle$ can
be expanded as a sum over paths within the $j$th phase as 
\begin{equation}
\langle p_{j+1}\vert (T_{v_{j}})^{h_{j}}\vert p_{j}\rangle = 
\sum_{\begin{array}{cc} \scriptstyle{paths \; q\; \; of } \\
\scriptstyle{length \; h_{j}+1} \end{array}}\langle
q_{h_{j}+1}\vert T_{v_{j}}\vert q_{h_{j}}\rangle \cdots \langle
q_{3} \vert T_{v_{j}}\vert q_{2}\rangle \langle q_{2}\vert
T_{h_{j}}\vert q_{1}\rangle \langle p_{j+1}\vert
q_{h_{j}+1}\rangle \langle p_{j}\vert q_{1}\rangle.
\label{Tinpathsum}
\end{equation}

This shows that paths within a phase can wander anywhere provided
they begin and end with states corresponding to the input and
output states for the phase. The path amplitudes are determined
by the properties of $T$ and are nonzero only if Eqs. \ref{Taext}
or \ref{Tcext} are satisfied.

These representations show that for implementation of a task as a
sequence of action and computation phases, it is necessary that
the initial and terminal states of completed phases have the
required properties. No requirements are given on intermediate
path states. The paths can wander anywhere in the overall system
state space. Of course the amplitude for any path depends on the
properties of $T$.

\section{Conclusion}
\label{Concl}

The example discused, of distance measurement by site counting, was kept
very simple  as a first example of a task as a decision tree of
computation and action phases.  No entanglements or basis changes were included.
More complex tasks that result in entanglements can be considered.  For
example Shor's or Grover's algorithms \cite{Shor,Grover} can be included in
tasks.  Also tasks that include decision trees of sequences of measurements
of noncommuting observables are possible.

As noted earlier a main reason for studying quantum robots and
their interactions with environments of quantum systems is that
these systems provide a well defined platform for investigation
of many interesting questions.  For example "What properties must
a quantum system have so that one can conclude that it is aware
of its environment, makes decisions, and has other properties of
intelligence?"  Answering such a question, even for models of
quantum robots plus environments as defined here, is not easy.
 It seems impossible without the framework of some model such
as that given in this paper. This is emphasized by the fact that
the only known examples of intelligent quantum systems are very
complex and contain the order of $10^{23}$ degrees of freedom.

It is also worthwhile to consider the following speculations. The
close connection between quantum computers and quantum robots
interacting with environments suggests that the class of all
possible physical experiments may be amenable to characterization
just as is done for the computable functions by the Church-Turing
hypothesis \cite{Turing,Deutsch85,Nielsen}.  That is there may be
a similar hypothesis for the class of physical experiments.

The description of tasks carried out by quantum robots (Section
\ref{QR}) lends support to this idea in that there may be an
equivalent Church Turing hypothesis for the collection of all
tasks that can be carried out.  The earlier work that
characterizes physical proceedures as collections of instructions
\cite{FouRan,Eks}, or state preparation and observation
proceedures as instruction booklets or programs for robots
\cite{BenEks} also supports this idea.  On the other hand much
work needs to be done to give a precise characterization of
physical experiments, if such is indeed possible.
  
\section*{Acknowledgements}
This work is supported by the U.S. Department of Energy, Nuclear 
Physics Division, under contract W-31-109-ENG-38.

\begin{center}
FIGURE CAPTIONS
\end{center}

Figure 1.  A Schematic Model of a Quantum Robot and its
Environment. The environment is a 3-D space lattice containing
various types of quantum systems (not shown). The quantum robot
QR consists of an on board Quantum Turing machine, a finite state
output system (o), and a control qubit (c). The on board QTM
consists of a finite closed lattice ${\cal L}_{2}$ of qubits and
a finite state head $h_{2}$ that moves on ${\cal L}{_2}$. The
location of a marker qubit (q) is shown. The position
$\underline{x} = (x,y,z)$ of the quantum robot (QR) on the
environment lattice is shown by an arrow. \\
\\
Figure 2.  A Schematic Model of a Quantum Robot for the Specific
Task on a 1-D Environment Space Lattice.  The particle (p) is not
shown. The other systems are as in Figure 1 except that the (m)
systems is expanded into an $N+2$ qubit lattice ${\cal L}_{3}$. 
The position of the quantum robot on the environment lattice is
shown with an arrow. \\
\\
Figure 3. Decision Tree for the Example Task.  Task process
motion is indicated by the arrows.  Circles represent action
phases.  Square boxes show relevant states of systems.  Permanent
storage and running memory are shown respectively by $st$ and
$d$. The boxes between adjacent action phase circles show what
occurs during a computation phase.  The lefthand column shows
task progress during the first search part.  The center column
with horizontal arrows shows what happens in a computation phase
when (p) is first located.  The righthand column shows task
progress during the return part.  The ballast activities that
occur when the task is complete are shown in the upper right. 
The actions $mr>$ and $ml>$ are nonhalting motion of
the quantum robot to the right and to the left.


\begin{thebibliography}{99}

\bibitem{Vedraletal}
V. Vedral, A. Barenco, and A. Ekert, Phys. Rev. A {\bf 54} 147
(1996)

\bibitem{Barencoetal}
A. Barenco, C. H. Bennett, R. Cleve, D. P. DiVincenzo, N.
Margolus, P. Shor, T. Sleator, J. A. Smolin, and H. Weinfurter,
Phys. Rev. A {\bf 52} 3457 (1995)

\bibitem{EkJo}
A. Ekert and R. Jozsa, Revs. Modern Phys. {\bf 68} 733 (1996)

\bibitem{Benioff8082}
P. Benioff, Jour. Stat. Phys. {\bf 22} 563 (1980); Phys. Rev.
Letters {\bf 48} 1581 (1982).

\bibitem{Benioff86}
P. Benioff, Ann. NY Acad. Sci. {\bf 480} 475 (1986)

\bibitem{Deutsch85}
D. Deutsch, Proc. Roy. Soc. (London) A {\bf 400} 997 (1985).

\bibitem{Deutsch89}
D. Deutsch, Proc. Roy. Soc. (London) A {\bf 425} 73 (1989)

\bibitem{BenioffQBE}
P. Benioff, Phys. Rev. A {\bf 54} 1106 (1996); Phys. Rev. Letters
{\bf 78} 590 (1997); to appear, Fortschrifte der Physik.

\bibitem{Shor}
P. Shor, in {\it Proceedings of the 35th Annual Symposium on the
Foundations of Computer Science}, edited by S. Goldwasser (IEEE
Computer Society, Los Alamitos, CA 1994), p. 124; Siam Jour.
Comput. {\bf 26}, 1481 (1997).

\bibitem{Grover}
L.K.Grover, in {\it Proceedings of 28th Annual ACM Symposium on
Theory of Computing} ACM Press New York 1996, p. 212; Phys. Rev.
Letters, {\bf 78} 325 (1997); G. Brassard, Science {\bf 275} 627
(1997).

\bibitem{Feynman82}
R. P. Feynman, International Jour. of Theoret. Phys. {\bf 21} 467
(1982)

\bibitem{Landauer}
R. Landauer, Physics Letters A {\bf 217} 188 (1996); Phil. Trans.
R. Soc. Lond. A {\bf 353} 367 (1995); Physics Today {bf 44} 23
(1991) May; IEEE Transactions on Electron Devices, {\bf 43} 1637
(1996).

\bibitem{Lf}
R. Laflamme, C. Miquel, J. P. Paz, and W. H. Zurek, Phys. Rev.
Letters {\bf 77} 198 (1996); E. Knill and R. Laflamme, Phys. Rev
A {\bf 55} 900 (1997); P. W. Shor, Phys. Rev A {\bf 52} R2493 (1995); D. P.
DiVincenzo and P. W. Shor, Phys. Rev. Letters {\bf 77} 3260 (1996).

\bibitem{KnLfZu}
E. Knill, R. Laflamme, and W. H. Zurek, Science, {\bf 279} 342
(1998).

\bibitem{PerZur}
A. Peres and W. Zurek, Amer. Jour. Phys. {\bf 50} 807 (1982).

\bibitem{Albert}
D. Albert, Physics Letters {\bf 98A} 249 (1983); Philosophy of
science {\bf 54} 577 (1987); {\it The Quantum Mechanics of Self-measurement} in {\bf Complexity, Entropy and the Physics of
Information}, proceedings of the 1988 workshop in Santa Fe, New
Mexico, 1989, W. Zurek, Ed.  Addison Wesely Publishing Co. 1990.

\bibitem{Breuer}
T. Breuer, Philos. Science {\bf 62} 197 (1995).

\bibitem{Peres}
Phys. Letters, {\bf A101} 249 (1984).

\bibitem{Penrose} 
R. Penrose, {\it The Emperor's New Mind}, Penguin Books, New
York, 1991.

\bibitem{Stapp}
H. P. Stapp, {\it Mind, Matter, and Quantum Mechanics}, Springer
Verlag, Berlin 1993.

\bibitem{Squires}
E. Squires, {\it Conscious Mind in the Physical World} IOP
Publishing, Bristol England, 1990

\bibitem{FouRan}
C. H. Randall and D. J. Foulis, Amer. Math. Monthly, {\bf 77} 363
(1970); D. J. Foulis and C. H. Randall, Jour. Math. Phys., {\bf
13} 1667 (1972).

\bibitem{BenEks}
P. Benioff and H. Ekstein, Phys. Rev. D {\bf 15} 3563, (1977);
Nuovo Cim. {\bf 40 B} 9 (1977).

\bibitem{HeMi}
C. D. Helon and G. J. Milburn, {\it Quantum Measurements with a
 Quantum Computer}, Los Alamos Archives preprint, quant-ph/9705014; S.
Schneider, H. M. Wiseman, W. J. Munro, and G. J. Milburn, {\it
Measurement and State Preparation via ion trap quantum computing}, Los
Alamos Archives preprint, quant-ph/9709042.

\bibitem{Lloyd}
S. Lloyd, Phys. Rev. A {\bf 56} 3374 (1997).

\bibitem{BBBV}
C. H. Bennett, E. Bernstein, Giles Brassard, and U. Vazirani, Siam Jour.
Computing, {\bf 26} 1510 (1997).

\bibitem{BBCM}
R. Beals, H. Buhrman, R. Cleve, and M. Mosca, {\it Tight Quantum Bounds by
Polynomials}, Los Alamos preprint archives quant-ph/9802049.

\bibitem{ref}
The author thanks the referee for pointing out the connection of oracle
quantum computing and Grover's algorithm to the present work.

\bibitem{BenSFA}
P. Benioff, Superlattices and Microstructures {\bf 23} 407 (1998).

\bibitem{Zur}
W. H. Zurek, Phys. Rev. D {\bf 24} 1516 (1981); {\bf 26} 1862
(1982); J. R. Anglin and W. H. Zurek, Phys. Rev. D {\bf 53} 7327
(1996).

\bibitem{Joo}
E. Joos and H. D. Zeh, Z. Phys. B {\bf 59}, 223 (1985).

\bibitem{Yao}
A. Yao, in {\it Proceedings of the 34th Annual Symposium on
Foundations of Computer Science} (IEEE Computer Society, Los
Alamitos, CA, 1993), pp. 352-361.

\bibitem{WooZur}
W. K. Wootters and W. H. Zurek, Nature {\bf 299}, 802 (1982); H.
P. Yuen, Physics Letters {\bf 113A}, 405 (1986); H. Barnum, C. M.
Caves, C. A. Fuchs, R. Jozsa, and B. Schumacher, Phys. Rev.
Letters {\bf 76} 2818 (1996); L. M. Duan and G. C. Duo, Los
Alamos Archives, preprint no. quant-ph/9705018.

\bibitem{Bennett}
C. H. Bennett, IBM Jour. Res. Dev. {bf 17}, 525 (1973).


\bibitem{Land1}
R. Landauer, {\it Zig-Zag Path to Understanding} in Proceedings
of the Workshop on Physics and Computation, PhysComp  94, Los
Alamitos: IEEE Computer Society Press, 1994.

\bibitem{BeVa}
E. Bernstein and U. Vazirani, in {\it Proceedings of the 1993 ACM
Symposium on Theory of Computing} (ACM, New York, 1993), pp 1-20.

\bibitem{Feynman}
R. P. Feynman, Optics News {\bf 11} 11 (1985); reprinted in
Foundations of Physics {\bf 16} 507 (1986).

\bibitem{FeHi}
R.P. Feynman and A. R. Hibbs, {\it Quantum Mechanics and Path
Integrals}, McGraw-Hill Book Co. New York 1965.

\bibitem{Turing}
A. Church, Am. Jour. Math. {\bf 58},345 (1936); A. M. Turing,
Proc. Lond. Math. Soc. 2 {\bf 42}, 230 (1936).

\bibitem{Nielsen}
M. A. Nielsen, Phys. Rev. Letters, {\bf 79} 2915 (1997); K
Svozil, {\it The Church-Turing thesis as a Guiding Principle for
Physics} Los Alamos Archives preprint quant-ph/9710052.

\bibitem{Eks}
H. Ekstein, Phys. Rev. {\bf 153}, 1397 (1967); {\bf 184}, 1315
(1969).

\end{thebibliography}
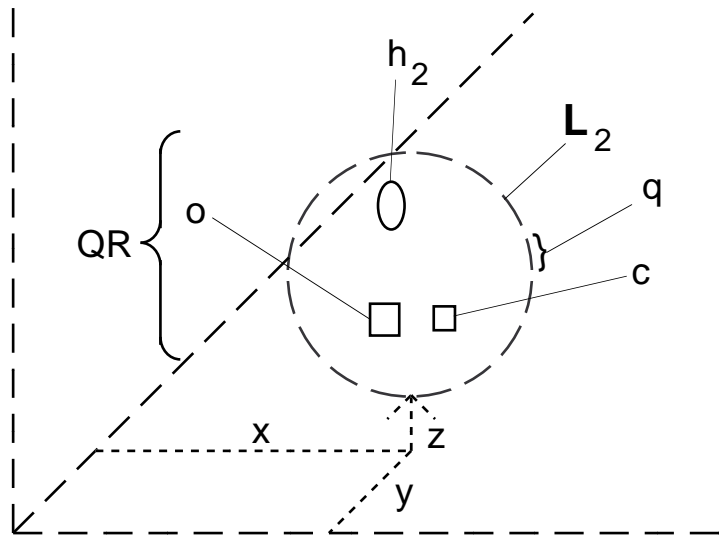
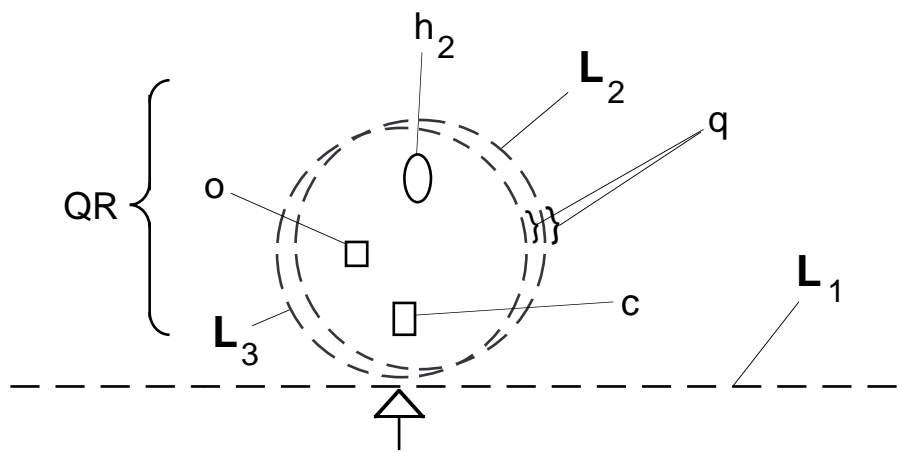
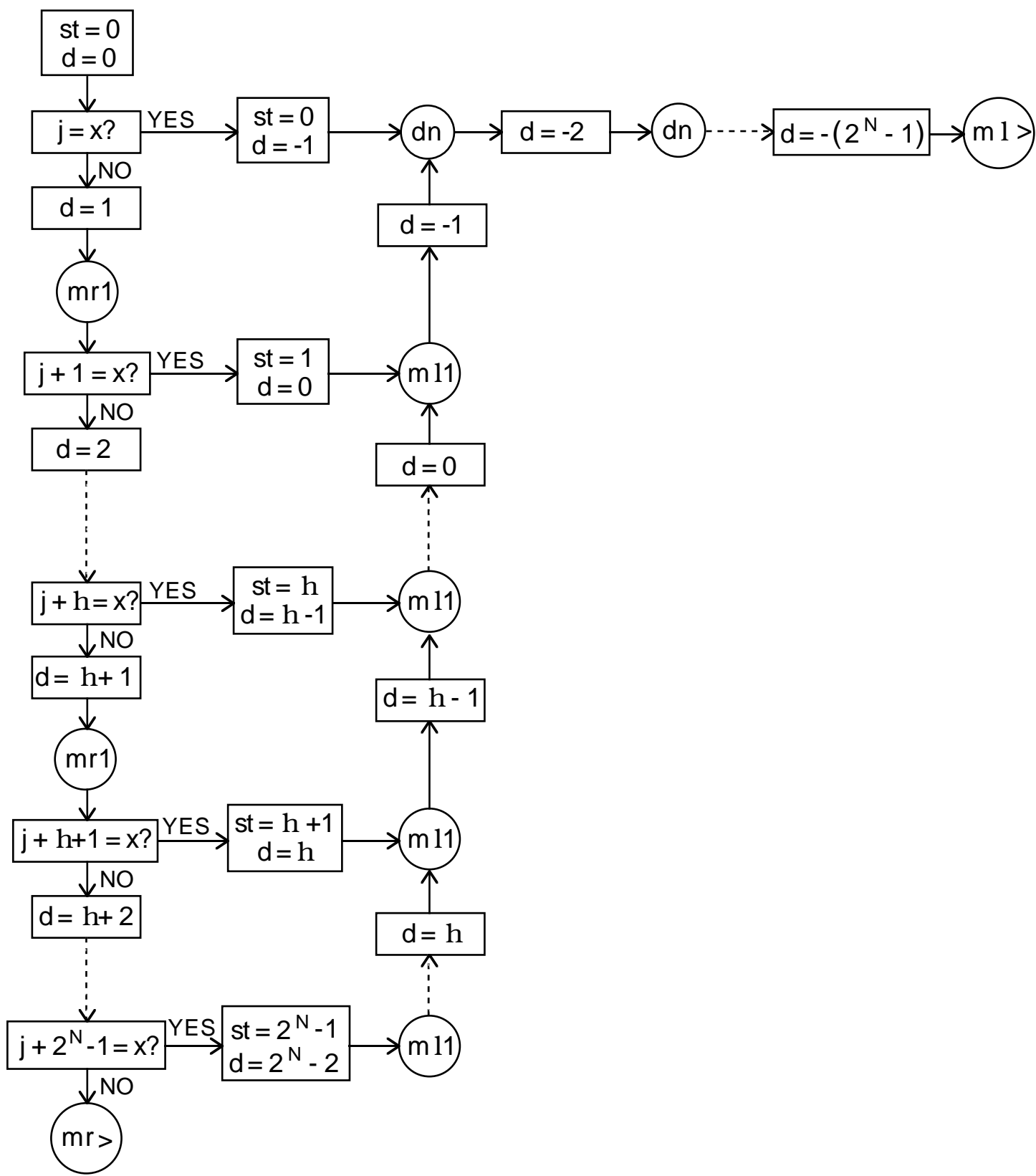
\end{document}